\documentclass[sigplan, anonymous=false, screen=true, natbib=true]{acmart}

\makeatletter
\renewcommand\@formatdoi[1]{\ignorespaces}
\makeatother

\makeatletter
\def\@copyrightspace{\relax}
\makeatother

\usepackage{url}

\usepackage[utf8]{inputenc}
\usepackage[english]{babel}
\usepackage{setspace}
\acmConference[Burlingame '21]{TinyML Research Symposium(TinyML Research Symposium’21)}{March 22, 2021}{Burlingame, CA}
\acmISBN{}

\usepackage{rotating}
\newcommand\tr[1]{\begin{turn}{90}\rlap{#1}\end{turn}}
\usepackage{booktabs}
\usepackage{varwidth}

\usepackage{footmisc}

\usepackage{float}
\usepackage{listings}
\usepackage{tablefootnote}
\usepackage{caption}
\usepackage{subcaption}

\usepackage{pifont}% http://ctan.org/pkg/pifont
\newcommand{\cmark}{\ding{51}}%
\newcommand{\xmark}{\ding{55}}%

\begin{document}
\title{Compiler Toolchains for Deep Learning Workloads on Embedded Platforms}
\author{Max Sponner}
\affiliation{
  \institution{Infineon Technologies \\Dresden GmbH \& Co. KG}
  \city{Dresden}
  \country{Germany}
}
\email{max.sponner(at)infineon.com}

\author{Bernd Waschneck}
\affiliation{
  \institution{Infineon Technologies \\Dresden GmbH \& Co. KG}
  \city{Dresden}
  \country{Germany}
}
\email{bernd.waschneck(at)infineon.com}

\author{Akash Kumar}
\affiliation{
  \institution{Dresden University of Technology}
  \city{Dresden}
  \country{Germany}
}
\email{akash.kumar(at)tu-dresden.de}

\begin{abstract}
As the usage of deep learning becomes increasingly popular in mobile and embedded solutions, it is necessary to 
convert the framework-specific network representations into executable code for these embedded platforms.\\
This paper consists of two parts: The first section is made up of a survey and benchmark of the available open source deep learning compiler toolchains, 
which focus on the capabilities and performance of the individual solutions in regard to targeting embedded devices and microcontrollers 
that are combined with a dedicated accelerator in a heterogeneous fashion.\\
The second part explores the implementation and evaluation of a compilation flow for such a heterogeneous device
 and reuses one of the existing toolchains to demonstrate the necessary steps for hardware developers that plan to build a software flow for their own hardware.
\end{abstract}



\maketitle

\section{Introduction}\noindent
As AI is moving closer to the edge, the limitations of the popular deep learning frameworks in regard to embedded platforms become apparent.
These platforms often employ Cortex-M CPUs or similar solutions and can operate on the (sub) milliwatt range for their power consumption.
Deep learning frameworks are mostly designed for the server and workstation use and incorporate many features that are not relevant for the inference on low power devices,
this prevents them from running on microcontrollers and other embedded solutions.\\
Due to this, the usage of deep learning models on embedded devices typically relies on the manual implementation of the previously trained networks.
The developers have to implement the required layer types, preferably using the vendor-provided math kernel libraries for the platform.
This process is labour intensive, error prone and can easily result in inferior performance due to missing or incorrectly executed optimizations.
In addition, the support of new platforms can result in extensive effort as the function kernels might need to be re-implemented for them.\\
The necessary modifications increase even further if dedicated deep learning accelerators are employed.
Due to their use of domain-specific instruction sets, which often utilize long pipelines that cover common neural network sub-patterns, these accelerators cannot easily be targeted by standard compiler toolchains.
Additional obstacles can be the coarse-granular accelerator instruction set architectures and the usage of custom data types, which can differ from the types used by the neural networks.\\
A possible solution to automate these tasks are deep learning compilers.
These toolchains operate similar to standard compilers, but introduce a number of important peculiarities:
Instead of handwritten source code, deep learning compilers process serialized trained neural network descriptions.
Additionally, they should be able to automatically employ the optimized math kernel libraries or alternative optimizations of the function kernels.\\
A benefit of employing domain-specific compilers is the option to introduce additional optimizations that target deep learning models.
Typically these are layer fusion or general network graph optimizations and quantization schemes.\\
Lastly, these toolchains should be able to target heterogeneous platforms and dedicated accelerators by employing runtimes on the target devices.
These runtimes take care of the scheduling and additional support operations that are necessary as well as the deserialization of the compiled networks.\\
This paper will start with an overview of the available optimizations and math kernel libraries for deep learning workloads on embedded platform, 
which should be incorporated by the compiler toolchains for optimal performance.
The following sections will cover a survey of the compiler features and the achieved performance on different embedded and low-power platforms, 
while the last part of the paper contains the implementation of a compilation flow for a new custom target.\\
The intended target audience of this paper are research teams that are looking for a software stack to support their own
deep learning hardware efforts, while achieving optimal performance and broad compatibility across the available frameworks.
%\begin{itemize}
%  \item difficult to deploy deep learning models to different platforms
%  \item each platform usually comes with own math kernel library and constraints
%  \item would require handcrafted reimplementation for each new platform
%  \item labor-intense and error prone
%  \item increases with the introduction of dedicated deep learning accelerators
%  \item usually come with domain-specific instruction set, utilizing longer pipelines and reduced control flow
%  \item unable to be targeted by standard compilers
%  \item solution: deep learning compiler
%  \item similar to standard compilers, but number of important differences:
%  \item operate on serialized trained models, generated by deep learning frameworks
%  \item should incorporate platform-specific math kernel libraries or similar optimizations automatically
%  \item employ additional domain-specific optimizations that make use of resilience to changes of neural networks
%  \item should be able to target heterogeneous platforms, incorporating the accelerator for its supported functions, but utilizes CPU for remaining subgraphs od models
%\end{itemize}

\section{Related Work}\noindent
Two recent studies focus on the employed techniques and the use with FPGA platforms \citep{xing2019depth}, 
as well as an in-depth overview over the different approaches for common problems of the available deep learning compilers \citep{li2020deep}. 
In contrast to these publications, this work focuses more on embedded platforms.
In addition to the survey and benchmark a compilation toolchain based on TVM has been implemented to target a heterogeneous platform.
This was done to demonstrate the steps that are currently required to support custom accelerators with their own software stack.

\subsection{Deep Learning Optimizations}\noindent
Deep Learning toolchains can employ a multitude of domain-specific optimizations in addition to standard compiler strategies.
These include weight pruning, which cuts redundant and unnecessary weights from the networks to reduce their size and - depending on the implementation - the compute workload.
While a wide range of pruning algorithms exist, none of them are currently employed by deep learning compilers \citep{han2015deep, lecun1990optimal, blalock2020state, wang2019structured}.\\
In contrast, quantization schemes are utilized by most deep learning compilers.
This optimization converts the floating point representations for weights, intermediate results and outputs into smaller fixed-point formats, while mostly keeping the accuracy of the original network.
This enables the network to take up less storage and reduces its bandwidth requirements during the execution as well as the computational intensity, if optimized function kernels for the quantized representations exist \citep{quant_qualcomm, quant_google, quant_google2, quant_facebook, petewarden_quant}.\\
Additional strategies include optimizations on the neural network graph like layer fusion, dead node elimination and others \citep{TF-graph-opt, TF2-graph-opt}.

\subsection{Math Kernel Libraries}\noindent
Math kernel libraries are a way to provide developers with optimized function kernels for common operations, increasing the efficiency of software solutions that employ them, while decreasing the redundancy of implementations.
They are typically platform-specific and provided by the device-vendors\footnote{Efforts for device-independent solutions exist as well, but have not found the same rate of adaption, e.g. XNNPack \citep{XNNPack}}.
%Table \ref{tab:mkl} presents an overview of the supported platforms per library. 
These libraries differ largely in their implementation strategies and offered functionality. 
While all libraries provide function kernels, their implementations follow different approaches:
ARM CMSIS-NN \citep{CMSIS-NN} mostly resorts to a low number of kernels that deliver consistent performance across the majority of the configuration space. 
In contrast, Intel's oneDNN\footnote{previously known as MKL-DNN} \citep{oneDNN_git} library implements most operations with multiple different strategies, 
based on the available instruction set extensions, data types and the configuration of the currently executed layer. 
The final selection of the function kernel takes places at runtime\footnote{only for its x86\_64 CPU backend} to achieve the best performance under the current circumstances.
While this strategy can be able to achieve better performance in certain cases, it requires much more maintenance and implementation effort compared to the more generalized function kernels.

\subsection{Deep Learning Accelerators}\noindent
In recent years plenty of deep learning accelerators emerged in commercial and research applications.
Commercial accelerators for embedded platforms include the NVDLA from Nvidia 
\footnote{The NVDLA is available as an open source project, which provides the hardware description and software stack, but is also contained in several of Nvidia's own products like the Jetson Xavier NX, using an alternative closed-source software stack} \citep{nvdla}, Google's EdgeTPU \citep{EdgeTPU} as well as ARM's Ethos-U NPUs \citep{EthosU}.
Most of these solutions employ custom compilation toolchains that are limited to support of only one deep learning framework for its input formats.\\
Research platforms include the Eyeriss accelerators (v1, v2) \citep{eyeriss_v1, eyeriss_v2}, VTA \citep{vta} as well as a many FPGA-based solutions \citep{fpga_rev}.
These typically do not focus on the software toolchain and explore novel architecture approaches instead.%\\
%FPGA deep learning high-level synthesis frameworks are also mentioned here as they represent a mixture of deep learning compilers and accelerators.
%These frameworks synthesize FPGA overlays from the deep learning models that are able to execute them \citep{fpga_toolflows}.

\section{Survey \& Benchmark}\noindent
The survey section of the paper covers open source projects that are still in development.
Its focus lies on the inference of deep learning models on embedded hardware.
The support for the training step will not be evaluated as it is uncommon to execute it on the embedded device itself.\\
The evaluated deep learning compilers are TensorFlow Lite (TFLite) \citep{tflite}, TensorFlow XLA (TF XLA) \citep{xla2019xla}, Glow \citep{glow}, TVM \citep{tvm}, ONNC \citep{onnc} and nGraph \citep{ngraph}, which has been tested as part of Intel's openVINO toolkit.\\
As µTVM was in an early stage at the time of testing, it has not been evaluated in this paper.

%\begin{itemize}
%  \item evaluated toolchains must be still in development
%  \item open source
%  \item only inference functionality has been included
%  \item evaluated compilers TF Lite, TF XLA, Glow, TVM, ONNC, openVINO/nGraph
%\end{itemize}

\subsection{Common Strategies}\noindent
All evaluated toolchains follow the typical compiler structure.
The frontend converts the serialized pretrained models into a high-level intermediate representation (IR).
Most toolchains utilize a two-level IR: The high-level IR is a graph-level representation of the compiled model and the low-level IR describes the operations on the tensor level.
The graph-level IR is typically used for mostly target-independent optimizations and operator fusion.
The tensor-level IR is used by the backend to optimize the individual operations.\\
One exception is TFLite, which does not perform target-dependent optimizations at the compilation stage and only uses a graph-level representation.
Instead, its compiler (called TFLite converter) generates a graph-level representation that does not contain execution details, as the device-specific function kernels are part of its runtime. 
This allows for a better portability of the compiler output across devices, but prevents more target-specific optimizations at the offline compilation stage.\\
The majority of the evaluated toolchains employs a runtime\footnote{The only exception that never uses a runtime is TensorFlow XLA, while Glow's AOT flow for CPUs does not require a runtime, it is deployed on other platforms}, which needs to be present on the target device to execute the compiled neural network.
The functionality of the runtime differs between projects. All runtimes provide support infrastructure to unpack the compiled neural networks and an API for the integration into the user program.
Solutions like TFLite and ONNC deliver the operation execution strategies for the platform as part of the runtime. 
Glow and TVM utilize the runtime for heterogeneous targets and in addition TVM requires the runtime for profiling during the auto-tuning step.
TVM delivers function kernels that have been generated by its auto-tuner alongside the model description and weights to the runtime.\\
The main difference between the evaluated projects is the provisioning of the function kernels.
Most solutions utilize handcrafted implementations that integrate math kernel libraries.
This requires maintenance and updating of implementations for each layer type across all supported platforms\footnote{TFLite, ONNC and others employ this strategy}.
To circumvent these limitations, TVM employs an Auto-Tuning solutions which tries to find the best suited function kernels by using an 
AI-guided flow that incorporates measurements and estimations of execution times on the real target.
Glow bypasses all of these concerns by reusing the same generalized function kernels across all targets, where its target-dependent optimizations are only applied by the LLVM backend for the selected target.

%\begin{itemize}
%  \item frontend is able to compile different deep learning container format
%  \item additionally device-independent optimizations, mostly at graph-level
%  \item quantization should be employed
%  \item two-level IR, graph-level and tensor-level
%  \item backend performs mostly device-dependent optimizations
%  \item backend can employ generalized kernels, math kernel libs or auto-tuning
%  \item runtime approaches are common, especially for heterogeneous platforms
%  \item runtime allows for better portability of compiler output, but moves some of the workload to online step
%\end{itemize}

\subsection{User Accessibility}\noindent
User Accessibility mostly depends on the user interface of the offline compiler stage, the integration of the compiler output into the target application and the supported input formats.\\
For the supported frameworks and formats, ONNX\citep{onnx} is the most important, as it is an industry standard and converts from most frameworks that exist. 
See table \ref{tab:formats} for an overview of the supported formats and frameworks of each compiler toolchain.\\
\begin{table}[ht]
 \begin{center}
 \caption[DL Compiler Frontends \& Supported Hardware]{Overview of the supported deep learning framework formats and target hardware platforms.}
 \label{tab:formats}
\bgroup
\def\arraystretch{1.1}
%\resizebox{0.85\linewidth}{!}{
\begin{tabular}{| l | c | c | c | c | c | c |}
  \hline
                                                                            &                    &                    &                    &                    &                    &                          \\
                                                                            &                    &                    &                    &                    &                    &                          \\
                                                                            &                    &                    &                    &                    &                    &                          \\
                                                                            &                    &                    &                    &                    &                    &                          \\
  \tr{                      }                                            & \tr{TVM}     & \tr{TF Lite} & \tr{TF XLA} & \tr{Glow}    & \tr{ONNC}  & \tr{openVINO}  \\
  \hline \hline
  ONNX\citep{onnx}                                               & \cmark & \xmark & \xmark & \cmark & \cmark & \cmark                     \\
  \hline
  TensorFlow\citep{tensorflow2015-whitepaper}  & \cmark & \cmark & \cmark & \xmark & \xmark & \cmark                     \\
  \hline
  TensorFlow Lite flatbuffer                                  & \cmark & \cmark & \xmark & \cmark & \xmark & \xmark                     \\
  \hline
  PyTorch\citep{paszke2019pytorch}                     & \cmark & \xmark & \xmark & \xmark & \xmark & \xmark                     \\
  \hline
  MXNet\citep{chen2015mxnet}                             & \cmark & \xmark & \xmark & \xmark & \xmark & \cmark                     \\
  \hline
  Caffe\citep{markham2017caffe2}                       & \cmark & \xmark & \xmark & \cmark  & \xmark & \cmark                     \\
  \hline
  Keras\citep{chollet2015keras}                             & \cmark & \cmark & \cmark & \xmark  & \xmark & \xmark                     \\
  \hline \hline
    x86\_64                                         & \cmark                 & \cmark                               & \cmark                 & \cmark & \cmark & \cmark                     \\
  \hline
  Cortex-A                                       & \cmark                 & \cmark                               & ?\tablefootnote{inconclusive data}                           & \cmark & \cmark & \xmark                     \\
  \hline
  Cortex-M                                       & \xmark\tablefootnote{µTVM was not ready at the time of testing}                 & \cmark                               & ?\footnotemark[7]                           & \cmark & \cmark & \xmark                     \\
  \hline
  GPU (CUDA)                                 & \cmark                 & \xmark                               & \cmark                 & \xmark & \xmark & \xmark                     \\
  \hline
  GPU (OpenCL)                              & \cmark                  & \cmark                               & \cmark                 & \cmark & \xmark & \cmark                     \\
%  \hline
%  Bare-Metal                                    & \xmark                 & \cmark                               & ?\footnotemark[7]                         & \cmark & \cmark & \xmark                     \\
  \hline
  Deep Learning Accelerator                                              & \cmark                   & \cmark                              & \cmark                 & \cmark & \cmark & \cmark                     \\
  \hline
\end{tabular}%}
\egroup
\end{center}
\end{table}\noindent
All compilers either support a command-line interface, like traditional compiler toolchains, or the use through a Python API, which allows for the integration in the original training script of the deep learning model.
One exception is Intel's openVINO that provides an additional graphical user interface through a web-interface \citep{intel_openVINO}.
This enables more direct feedback to the developer on the impact of different optimizations on the overall model performance.\\
For the integration into the user application, all toolchains provide a C or C++ API. TVM and TFLite provide an additional Python interface through their standard runtimes\footnote{TFLite provides an additional runtime for microcontrollers, which does not come with a Python API}.\\

\subsection{Supported Platforms \& Extensibility}\noindent
As the support for low-power embedded devices\footnote{e.g. ARM Cortex-M and similar} in the currently available deep learning compilers is still limited,
additional platforms have been used during the evaluation to allow for a more complete performance comparison.\\
The range of supported platforms varies between the evaluated toolchains.
In addition, the level of optimization for the supported platforms fluctuates widely.
One such example is TFLite's support of the x86\_64 platform:
While its runtime can be compiled for it, the function kernels are not optimized, resulting in worse performance compared to other platforms or compilers.\\
For TF XLA no conclusive information about its support for different architectures could be found, 
as the official documentation and publications contradict each other \citep{tf_xla_doc, leary2017xla}.
The support for bare-metal use cases and embedded processors is much less common, as only TFLite, ONNC and Glow are able to target Cortex-M CPUs.
For an overview of the supported platforms see table \ref{tab:formats}.\\
While all toolchains include some kind of support for heterogeneous platforms, the implementations differ between them.
The most complete solution has been provided by TVM in its Bring-Your-Own-Codegen (BYOC) flow \citep{tvm_byoc}, 
which allows developers to target new libraries and accelerators from TVM's high-level IR.
It does not only provide an API to include new backends, it also supports the developer by implementing solutions for common problems, 
like the CPU fallback for unsupported operations, an infrastructure for custom operator fusion rules and the option to change the data layout of tensors.
Most other toolchains only supply a simple API to access values and layer configurations and require the developer to reimplement many of these common support tasks.\\
A stark contrast to TVM is TFLite. 
Its compilation stage does not provide an interface for the inclusion of additional target-specific tasks and optimizations.
New platforms are targeted by porting the runtime to them and deploying optimized function kernels with it.
As this flow only allows the targeting of general purpose hardware, its support for the currently available accelerators has been realized by additional tools.
These modify the flatbuffer file, which has been generated by the offline compilation stage, before it can be executed on the 
heterogeneous platform \citep{tflite_vela, EdgeTPU_compiler}. 
This approach breaks the portability of TFLite and requires additional work to keep these tools compatible with the current scheme of the TFLite flatbuffer files.
Some compiler toolchains like Glow, which reuse LLVM's backends\footnote{for its AOT flow} can easily target new architectures, if a LLVM backend already exists.
In that case Glow's ahead-of-time (AOT) compilation mode can be reused, if other targets need to be supported, a separate runtime can be used and the AOT flow can no longer be utilized.

\subsection{Features}\noindent
For the embedded use case, the AOT compilation and quantization support are the most important compiler features.
For the optimal execution on the target devices, the toolchains should be able to incorporate math kernel libraries or auto-tuning to provide optimal function kernels for the execution.
Features like support for the backpropagation and training steps are not as important for embedded devices (yet) and are only supported by TF XLA, Glow and openVINO as they primarily target HPC and cloud applications.\\
For the optimization of the layer execution TFLite, ONNC and openVINO rely on math kernel libraries, while Glow utilizes the same set of generalized function kernels across all targets, only utilizing the LLVM backend optimization steps.
TVM is the only evaluated toolchain that employs an auto-tuning process to find good performing function kernels automatically, but can also exploit third party libraries through the BYOC flow.\\
All toolchains - with the exception of TVM - implement only a limited number of static quantization schemes with fixed bit-widths for intermediates and weights.
This is a limitation for the targeting of custom devices as they could employ alternative quantizations.
TVM has implemented a more flexible quantization system, that offers different sizes and requantization strategies.
However, it is more difficult to configure compared to the other solutions and did not achieve competitive accuracies in this benchmark.
%A feature with growing importance is the support for dynamic shapes 
%as they are commonly used for natural language processing tasks. Only TVM, openVINO and TFLite\footnote{only since release 2.3 \citep{tf_v2.3}} are able 
%to compile networks that use dynamic shapes.

%\begin{itemize}
%  \item for embedded AOT, Quantization and Auto-Tuning are important
%  \item most compilers support these
%  \item auto-tuning is an alternative to math kernel library based handwritten function kernels
%  \item uses AI-guided process to find good implementations on target platform
%  \item JIT, Training, Automatic Differentiation are not important for embedded use case (yet)
%  \item dynamic shape grows in importance due to NLP tasks, no support in many toolchains yet
%  \item quantization might be a problem for accelerators, as most compilers only support one or two hardcoded variants
%  \item only TVM offers flexible solution, but hard to configure to get acceptable accuracy
%\end{itemize}
\subsection{Performance}\noindent
The performance was evaluated on an ARM Cortex-M55 fast model\footnote{as no hardware was available at the time of testing}, 
an Cortex-A72\footnote{using a Raspberry Pi 4 with 4\,GB of system memory} and a Intel Celeron J1900\footnote{using 8\,GB of DDR3 system memory}. 
The Cortex-A and Intel platform have been selected, to allow for a more complete performance overview due to the limited support of Cortex-M in the tested toolchains.
This also allowed for a direct comparison to the standard TensorFlow framework on these two platforms.\\
All of these platforms provide a SIMD vector extension and have been tested with the same simple MNIST test network, consisting of convolutional, fully connected, ReLU and maximum pooling layers.
The batch size has been set to one and the final activation function has been removed after training. These are common optimizations for embedded applications, as it reduces the jitter that is introduced by the 
predicition as well as the latency, as the final activation does not change the classification result.\\
%\begin{figure}[ht]
%    \centering
%    \includegraphics[width=\linewidth]{graphics/M55/instruction_count.pdf}
%  \caption{The instruction counts for a single inference that have been estimated by the Cortex-M55 fast model.}
%  \label{m55_ic}
%\end{figure}\noindent

\begin{figure}[ht]
    \centering
    \includegraphics[width=1.0\linewidth]{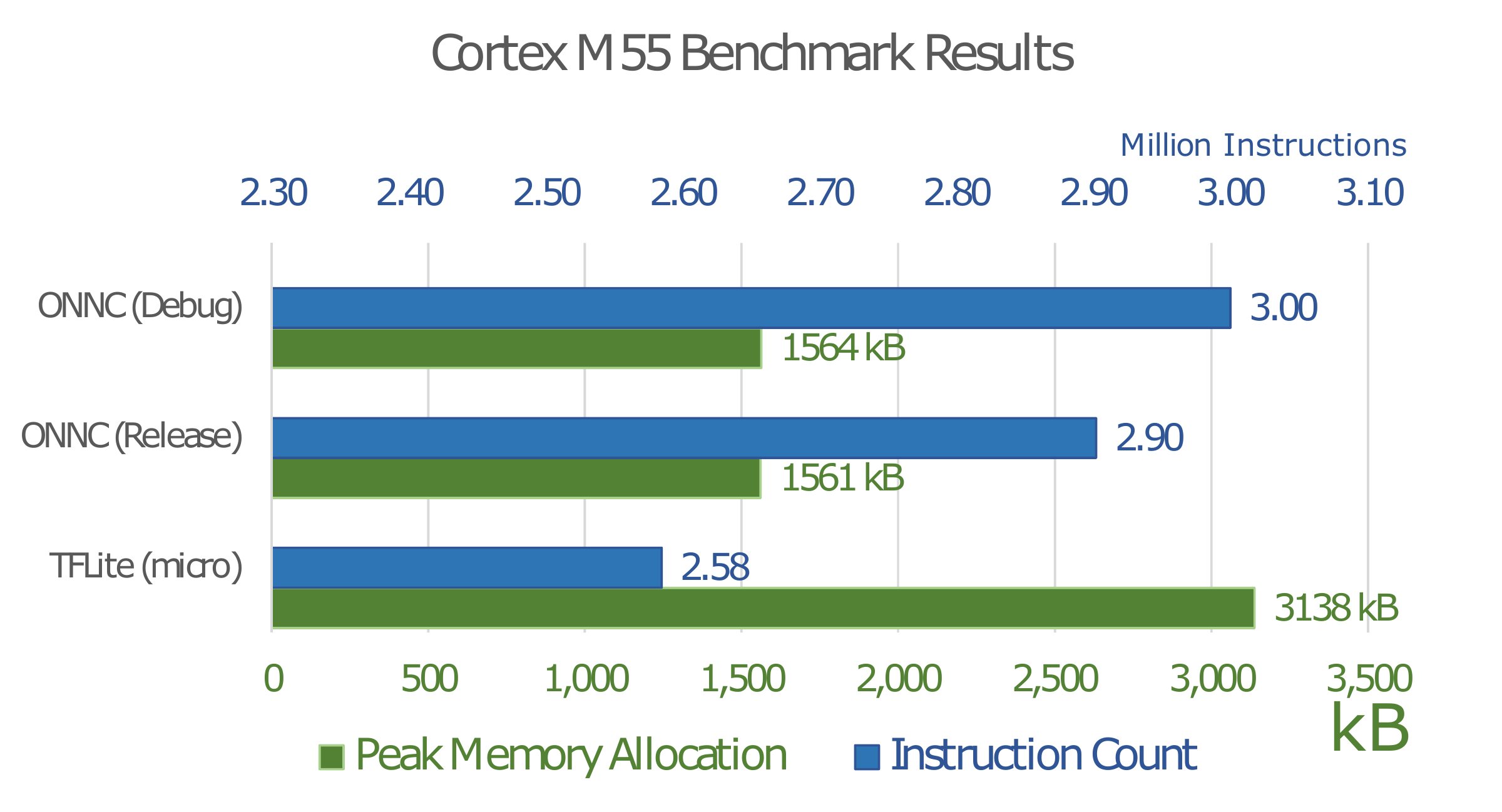}
  \caption{Instruction Counts and Peak Memory Allocation for the execution of the benchmark model on the Cortex M55.}
  \Description{Bar chart: This diagram shows the instruction count and peak memory allocation of the benchmark network after being compiled with TFLite and ONNC.TFLite's output required 2.58 million instructions and allocated 3138 kB of memory at its peak. ONNC was tested with the Debug and Release configuration of the compiler and required 3 million instructions with 1564 kB of memory (Debug) or 2.9 million instructions and 1561 kB for the Release configuration.}
  \label{m55_ic} \label{m55_mem}
\end{figure}
\begin{figure} 
  \centering
  \includegraphics[width=1.0\linewidth]{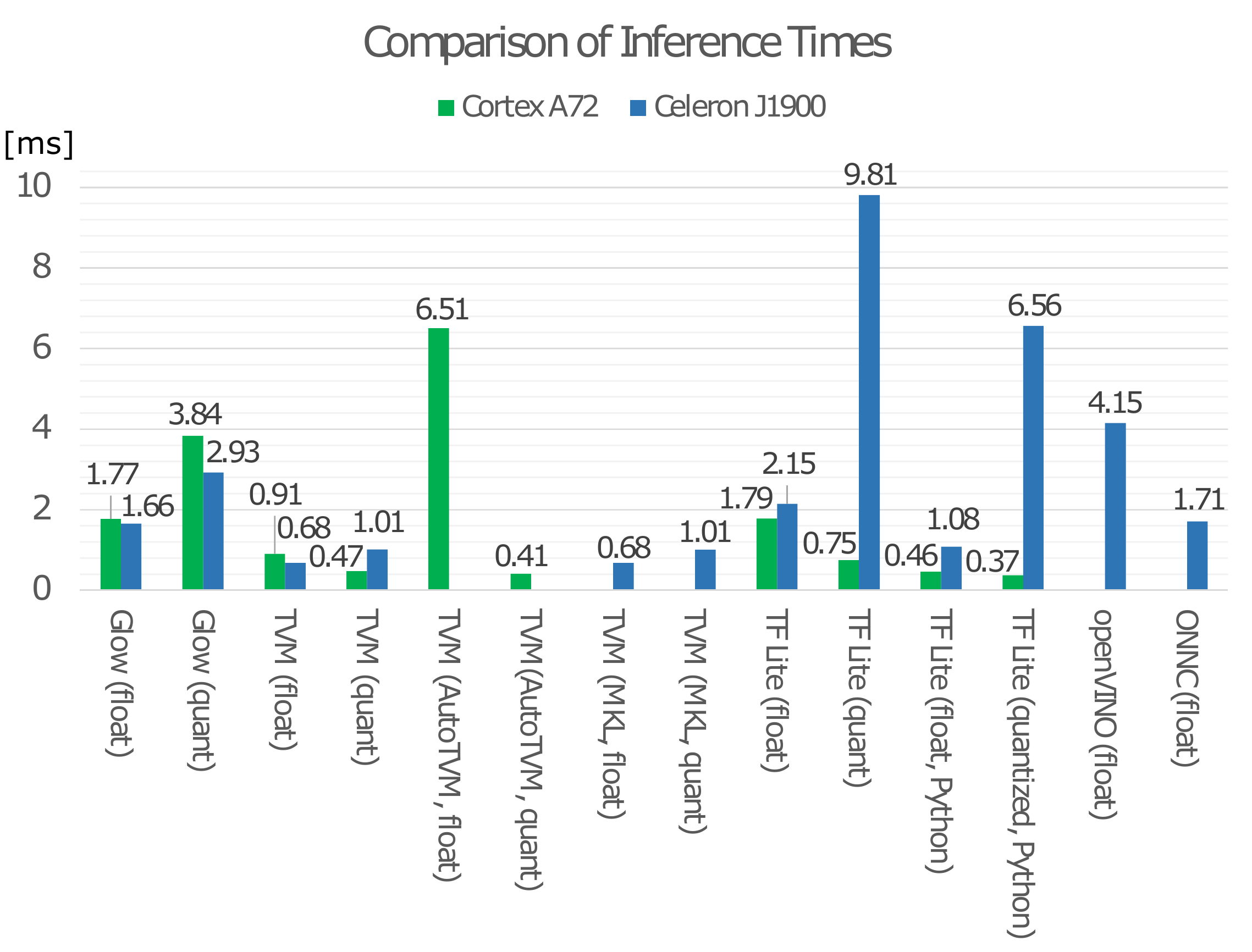}
  \caption{Comparison of the inference times on the ARM Cortex-A72 and Intel Celeron J1900 platforms.}
  \Description{Bar chart: Inference Times of tested toolchain outputs on Cortex-A72 and Intel Celeron J1900.
Glow with floating point data achieved 1.77 ms on the A72 and 1.66 ms on the J1900.
the quantized version took 3.84 ms on A72 and 2.93 on J1900.
TVM's float output took 0.91 ms (A72) and 0.68 ms (J1900).
The TVM quantized model took 0.47 ms (A72) or 1.01 ms (J1900).
The auto-tuned TVM model required 6.51 ms on the Cortex-A, while
the quantized version took 0.41 ms.
The MKL TVM model took 0.68 ms on the J-1900 for the floating point variant and
1.01 ms for the quantized network.
TFlite with its C-API and floats took 1.79 ms on A72 and 2.15 ms on the Celeron.
The quantized network using the C API required 0.75 ms on A72 and 9.81 ms on J1900.
For the Python API, the runtimes are 0.46 ms (float, A72), 1.08 ms (float, J1900)
and 0.37 ms (quantized, A72) and 6.56 ms (quantized, J1900).
Intel's openVino compiler achieved 4.15 ms and the ONNC 1.71 ms, both could only target the Intel platform.}
  \label{inference_time}
\end{figure}

\noindent
The Cortex-M55 could only be targeted by TFLite\footnote{using its micro-runtime} and ONNC\footnote{using a special Cortex-M version of the toolchain}.
As no hardware of the M55 is available yet, a instruction-level simulator has been used instead.
While Glow is able to target Cortex-M CPUs, it was not able to support the novel instruction set of the M55 (ARMv8.1M). 
%The available solutions had to be build using ARM's GCC compiler to target the platform. 
The testing showed that TFLite required less instructions to complete an inference run (2.6\,M instead of 3\,M instructions, see figure \ref{m55_ic}), while ONNC allocated significantly less memory (1.6\,MiB instead of 3\,MiB.
See figure \ref{m55_mem} for details).\\
%\begin{figure}[ht] 
%    \centering
%    \includegraphics[width=\linewidth]{graphics/M55/memory_allocation.pdf}
%    \caption{The peak allocations of RAM and ROM for the available toolchains during a single inference (batch size = 1).}
%  \label{m55_mem}
%\end{figure}\\
The next test platform was the Cortex-A72.
ONNC could not be compiled for it, as it relied on Intel MKL for its function kernels\footnote{While the CMake script for the standard runtime contained a parameter to disable the MKL integration, it could not be build when it was selected}.
Instead Glow, TVM and TFLite have been tested in addition to the standard TensorFlow Python runtime. An overview of the measured inference times can be seen at figure \ref{inference_time}.\\
The quantized TFLite version achieved the fastest inference speed with 0.37\,ms, followed by the quantized and auto-tuned TVM program, using its C runtime (0.41\,ms).
Glows fastest output achieved a mean inference time of 1.77\,ms by using floating-point representations.
Besides Glow, all compilers achieved faster inference times using quantized networks - suggesting that they employ optimized function kernels for quantized operations, while Glow uses its generalized standard kernels and wraps them into additional quantization and dequantization steps, which causes additional overhead.
The worst result by a compiler was achieved by TVM, for its floating-point auto-tuned solution, as it tried to apply x86 function templates to the ARM platform\footnote{it could not be determined, if it was caused by user error or by TVM, but it reoccurred over multiple tries and did not affect the quantized version}.
However, the slowest result of 6.51\,ms was still significantly faster than the use of the network in combination with the standard TensorFlow runtime - requiring 104.64\,ms for a single inference run.
This makes the slowest and incorrectly optimized compiled version 16 times faster, while the fastest compiled version achieved a speedup of 282.6 times.\\
\noindent
%\begin{figure*}[ht]
%\centering
%\setlength\fboxsep{0pt}
%\setlength\fboxrule{0.0pt}
%\fbox{\includegraphics[width=0.9\linewidth]{graphics/inference_times.pdf}}
%\caption{Comparison of the inference times on the ARM Cortex-A72 and Intel Celeron J1900 platforms.}
%\label{inference_time}
%\end{figure*}\noindent
The Intel Celeron CPU allowed for the additional testing of nGraph\footnote{as part of openVINO} and ONNC's standard flow\footnote{besides its Cortex-M version}.
See figure \ref{inference_time} for the inference time results of the platform.
In comparison to the Cortex-A results the ranking of the toolchains by their inference time changed, suggesting different levels of optimizations across the supported target devices for some deep learning toolchains.
In addition, TVM was tested with the Intel MKL BYOC-based backend instead of its auto-tuning flow.
This backend is not primarily optimized for performance as it is a demo for the BYOC functionality and was used to estimate the overhead which results from it.
For the Celeron J1900, the floating-point versions of the compiled networks achieved faster inference speeds across all toolchains.
This suggests either a lack of optimized kernels or a better implementation of the floating-point components of the hardware itself.
The fastest results have been achieved by TVM with 0.68\,ms (FP) and 1.01\,ms (quantized).
TVM did not show a significant difference between the standard and the BYOC flow results, which implies that the overhead of the BYOC flow is minimal.
The next best results were achieved by TFLite's floating point network (1.08\,ms, using the Python API), Glow (also floating point, 1.66\,ms) and ONNC (1.71\,ms).
openVINO's compiled program did require 4.15\,ms for a single inference, which made it the slowest floating point version out of the tested compiled networks.
It was not able to quantize the network, as that is only supported on newer Intel CPU generations.
Only TFLite's quantized networks took more time than openVINO to complete their inference run.\\ %TODO: extend, not complete yet
In addition to the inference times, the peak memory allocations have been measured.
The measured results varied by two orders of magnitude between the toolchains.
Glow's compiled networks required 10\,MiB of system memory at peak, followed by TVM with 21\,MiB to 26\,MiB.
As the higher allocations have been measured for the MKL-BYOC variant, it suggests, that the BYOC flow requires some memory overhead compared to the standard flow during the execution.
TFLite required 14\,MiB for a quantized version of the network utilizing only its C-API, which took significantly longer than the other results for a inference.
The same configuration, but with a floating point version of the network allocated 238\,MiB which is more than the expected increase by four times\footnote{as 8-bit integers are 4 times smaller than 32-bit floating point values}.
ONNC could only be tested with a floating point network as its open source standard branch does not support quantization.
Its peak memory allocation of 51\,MiB is more in line with the expected memory allocation.
openVINO's implementation allocated 489\,MiB of memory during the inference, only TFLite's Python runtime used more memory with 896\,MiB (quantized) or 1,248\,MiB (floating point).
While this values are significantly higher than the results of the other toolchains, they are still an improvement in comparison to the standard TensorFlow framework that allocated up to 2.3\,GiB.
\begin{figure} 
  \centering
  \includegraphics[width=1.0\linewidth]{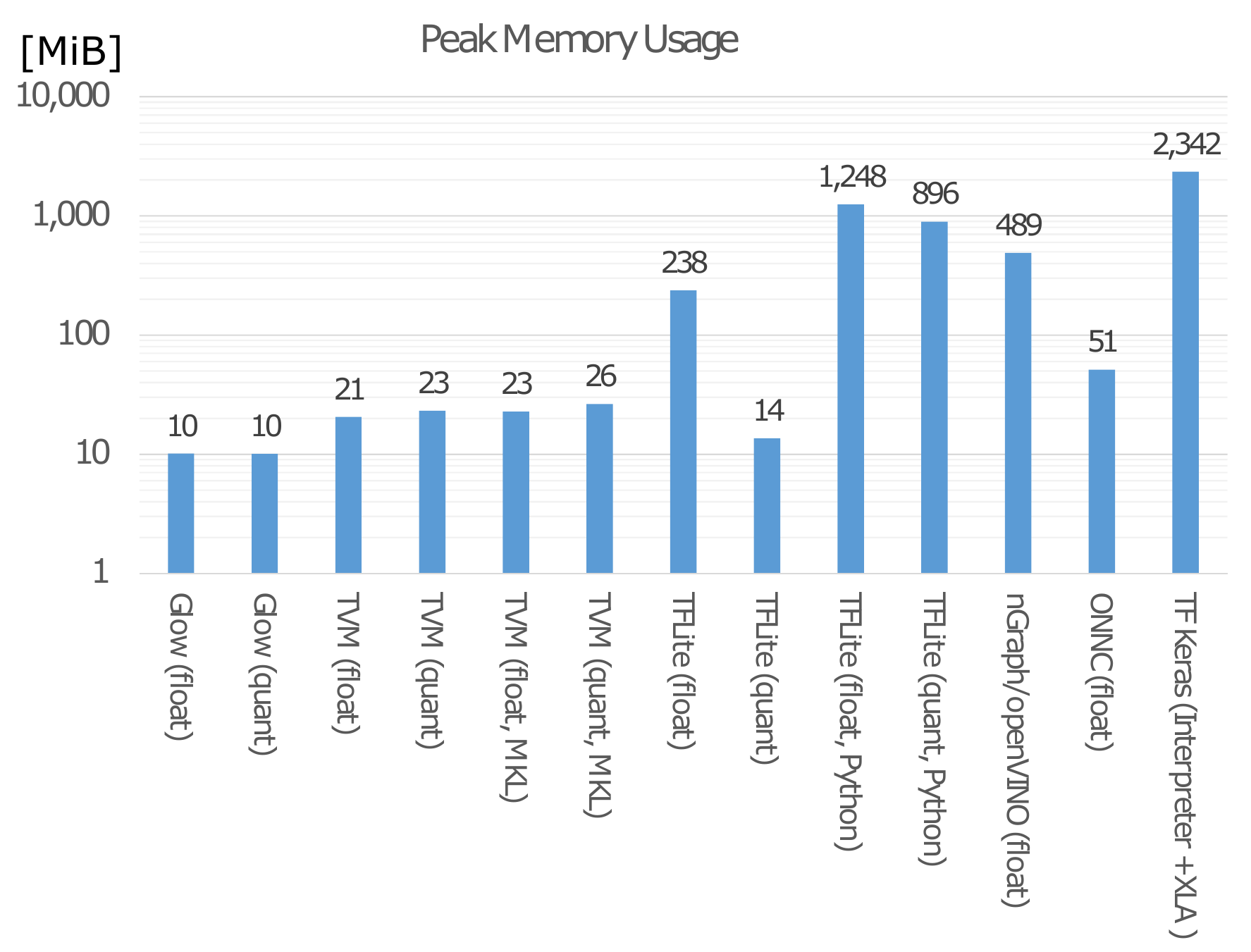}
\caption{Comparison of the peak memory allocation on the Intel Celeron J1900 platform.}
\Description{Bar chart: Peak Memory Allocation (MiB) on the Celeron J1900 platform.
Glow (float): 10MiB
Glow (quantized): 10 MiB
TVM (float): 21 MiB
TVM (quantized): 23 MiB
TVM (float, Intel MKL): 23 MiB
TVM (quantized, Intel MKL): 26 MiB
TFLite (float): 238 MiB
TFLite (quant): 14 MiB
TFLite (float, Python API): 1,248 MiB
TFLite (quant, Python API: 896 MiB
openVINO (float): 489 MiB
ONNC (float): 51 MiB
TensorFlow framework using Keras API: 2,342 MiB
}
\label{inference_mem}
\end{figure}\noindent
%\begin{figure*}[ht]
%\centering
%\setlength\fboxsep{0pt}
%\setlength\fboxrule{0.0pt}
%\fbox{\includegraphics[width=0.9\linewidth]{graphics/J1900/memory_no_TF.pdf}}
%\caption{Comparison of the peak memory allocation on the Intel Celeron J1900 platform.}
%\label{inference_mem}
%\end{figure*}\noindent
The prediction accuracy for the compiled networks stayed mostly the same, even for the quantized variants, with the exception of TVM.
It only reached an accuracy of around 50\,\%, which might have been user error due to its configurable quantization scheme and the usage of a global quantization scheme.\\ %following: summary of benchmark
The benchmark has shown that the available deep learning compilers are all able to deliver significantly better performance compared
to the use of the standard TensorFlow framework.
Additionally, they allowed for fast deployment of models on the evaluated platforms, without the need for handcrafted layer implementations.\\
TVM was able to deliver the best inference speeds on the larger test devices; nonetheless, its accuracy was limited by the quantization flow, which
was the only flexible quantization system out of the tested toolchains. 
While its frontend offered support for the majority of neural network serialization formats, 
the backend and runtime are unable to target the Cortex-M platform yet\footnote{However, a mico-runtime is currently in development, but not production ready yet}.
Its newly introduced BYOC flow allows developers to target heterogeneous platforms, while reusing the frontend for different input formats
and the standard components for the execution of unsupported operations on the system CPU.\\
TFLite was able to target the Cortex-M55 platform and achieved a similar performance to TVM on the Cortex-A system for the measured inference times.
Nevertheless, it allocated significantly more memory for its execution and does not offer a simple flow to target heterogeneous systems that include a dedicated accelerator. 
Additionally, users are limited to the capabilities and formats of the TensorFlow ecosystem.\\
Glow's support for the ONNX standard makes it compatible with most deep learning frameworks and its quantization flow achieves similar accuracies
to TFLite. While it achieved the lowest peak memory allocation out of the tested solutions, the use of generalized kernels lead to slower inference times compared
to the other platforms. For the targeting of embedded microcontrollers, the inclusion of CMSIS-NN could result in a significant performance improvement.\\
ONNC was able to target the Cortex-M55 system and delivered competitive performance to TFLite's micro-runtime. However,
the separation of the quantization tool into a commercial product, outside of the open-source project and the lack of support for newer ONNX subsets
might limit its use in future endeavors.\\
Intel's openVINO is only able to target the Celeron platform, but did not achieve a competitive performance result on this platform.
Its limitation to x86 CPU's makes it unable to target any kind of low-powered embedded platform. However, its graphical user interface (called workbench)
made the optimization and compilation process more transparent for the user, which could be a helpful feature for other toolchains as well.\\
While the number of supported layer functions is important for the user, it is difficult to compare these toolchains based on a single snapshot of
their development stages, due to constant updating of most of them.\\
This survey has shown that the support for embedded platforms in the current deep learning compiler toolchains is still at an early stage
as only a small subset is able to target platforms like the Cortex-M55.

\section{Implementation}\noindent
For the implementation, an abstract accelerator has been defined and an instruction set simulator was implemented\footnote{the simulator was written to only target 
the desired ISA and operates on the custom output files of the new compiler backend, which contain the initial memory layout and instructions for each subgraph.}. 
%This allowed for more flexibility in regards to its memory architecture, instruction set and register file.
The simulator was verified with TensorFlow and is able to estimate the hardware utilization and cycle count for input and output parallel execution strategies\footnote{input parallel: input channels are processed in parallel; output parallel: same for output channels}.\\
The simulated accelerator uses an instruction set that is similar in its capabilities to other solutions like Nvidia's NVDLA \citep{nvdla}.
It only supports signed integer formats for operations and the majority of them are limited to a length of eight bit for the individual values in their input tensors
while producing outputs with a length of either eight or 32\,bit.\\
For the software flow TVM was used due to its BYOC functionality.
This flow starts with the definition of annotation rules for supported nodes and patterns in TVM's graph-level IR\footnote{called Relay}.
These are then merged into subgraphs, which will be executed by the accelerator. These steps are handled by TVM's BYOC flow and did only require
the definition of supported graph patterns during the implementation of the new backend.
TVM manages the execution of unsupported operations on the CPU as well as the invocation of the subgraphs from the standard runtime.
After the annotation, the network graph is partitioned to separate the supported sections of the network into subgraphs.
These subgraphs are then passed on to the custom code generation, where they are converted into a JSON format for better portability across different instruction set variants.
The final generation of the accelerator command stream happens at runtime before the initial inference.
This allows to target different ISA variants with varying memory sizes using a single serialized file.
A custom runtime component executes the code generation and passes back a run function for each subgraph to the standard TVM graph runtime. During inference the subgraphs are
executed by the simulator through the invocation of these run functions.\\
Besides the quantization and data layout transformation functionality, which was provided by TVM, the memory planning for DMA operations between system and accelerator memory, the assembly generation, the configuration register file management and tiling for larger tensor sizes needed to be implemented by the custom runtime component. 
%The configuration register file was used, as the layer operations required too much parameters to fit into a single instruction and most parameters do not change in subsequent operations, especially when tiling is required.
The tiling was implemented by primarily splitting the workload along the output channel dimension.
%\begin{itemize}
%  \item keep it short
%  \item used TVM's BYOC flow
%  \item custom quantization
%  \item target was abstract accelerator with instruction set simulator
%  \item ISA similar to NVDLA capabilities, combined with exclusive memory and system memory
%  \item includes longer pipelines for convolutions + activation + requantization
%  \item tiling was implemented, to fit larger layers into limited device memory
%\end{itemize}
\section{Evaluation}\noindent \label{eval}
The correct functionality was initially tested with neural networks that only contained supported operations as single nodes and patterns.
Additional testing with the MNIST network from the performance benchmark revealed that the current TVM quantization flow inserts additional meta nodes into the graph.
These nodes prevent the merging of multiple compute layers into a single subgraph.
Due to this, the network was split into three subgraphs, which requires the system to move the intermediate results back to the shared system memory between the subgraph executions.
This resulted in reduced throughput and efficiency due to unnecessary bus transactions.
Otherwise, the custom backend worked as intended and generated the expected inference results.\\
For a more realistic test case Google's MobileNetV1 \citep{howard2017mobilenets} with the ImageNet dataset \citep{deng2009imagenet} has been used.
The additional batch normalization layers prevented the use of the larger convolutional pipelines, as they are located between the convolutional and activation function nodes.
This adds additional bus transactions between accelerator memory and compute units, which reduces the efficiency of real hardware.\\
The network was evaluated with three iterations of the accelerator and its toolchain:
\begin{itemize}
  \item Without support for depthwise convolutions: \\Depthwise convolutional layers are executed by the standard runtime on the CPU.
  This results in a lower share of the network to be accelerated by the target device. The estimated cycle counts can be 
  seen in figure \ref{fig:no_depth}.
  \item Software-only Implementation: \\ The depthwise convolutions are mapped to a the standard convolution instruction of the DLA.
  The ISA has not been changed. This allows for a larger share of the network to be accelerated, but the depthwise layers are
  constraint to a small utilization of the processing elements in the vector ALU.
  \item Full support: \\The ISA has been extended to provide a dedicated instruction for deptwhise convolutional layers.
  This instruction allows for a higher utilization, resulting in shorter execution times.
\end{itemize}
This was done to evaluate the flexibility of the coarse-grained ISA for the support of new layer types, as the deep learning research community develops new layer types at a very rapid pace, which makes it difficult to develop accelerators that can stay up-to-date for extended time periods without changing the hardware architecture.
A possible solution would be to update the software flow, to enable it to map new layers to existing instructions.
In the case of depthwise convolutions, this approach was represented by the second test scenario.
While the implementation was possible, it resulted in drastically increased cycle counts if compared to a native solution (compare figure \ref{fig:sw_depth} and \ref{fig:hw_depth}).
This was caused by the low utilization of the processing elements (PEs) as only one filter could be processed at a time.\\
Additionally, these scenarios were tested with different sizes of the accelerators exclusive memory and the vector ALU as well as input and output parallel mode for cycle and utilization estimations to evaluate the impact of operation tiling on the overall performance.
The evaluated memory sizes were 512\,KiB and 256\,MiB.
The last configuration does not require any tiling to take place, while the first is the smallest size which is supported by the implementing tiling methods for this network\footnote{For convolutional layers only a split along the output channel dimension was implemented, as the splitting along the rows and columns requires extensive effort to implement and validate all edge cases that can occur}.\\
As shown in figure \ref{fig:no_depth}, the output parallel hardware achieved lower cycle counts due to its higher utilization in the early layers of the network. However, this changed as soon
as tiling was required due to the limited amount of device memory. As the tiling splits the workload along the channel dimension of the output tensor, the parallelization opportunity for this architecture shrinks.
This results in a lower utilization, and a faster execution of input parallel strategies.
Additionally, it can be seen in figure \ref{fig:hw_depth}, which includes an additional 1024\,KiB configuration, that the cycle count does not scale linearly with the available memory.
Instead, a combination of larger vector ALU with 128\,processing elements (PEs) and 1024\,KiB of memory can be faster than a configuration with 64\,PEs and 256\,MiB of memory.
The reason is, that the 1024\,KiB configuration only requires tiling of the initial layers of the network, which allows it to compensate for the additional cycles in the later layers
where it can calculate twice as many results per cycle as the 64\,PE configuration with 256\,MiB of memory.
A benefit of the TVM BYOC flow is the ability to quickly evaluate the performance for different real-world network architectures across multiple hardware configurations during the hardware development.

\section{Conclusion}
This evaluation has shown that, while all evaluated toolchains deliver reasonable performance across different platforms, their support for low-powered embedded devices and heterogeneous solutions with dedicated accelerators is still at an early stage. 
Some are limited by the use of existing compiler backends, which prevent the targeting of dedicated hardware.
Another limitation is the use of static quantization schemes that do not offer an easy solution to adapt them for different hardware implementations.\\
Additionally, while it was possible to target a novel accelerator using TVM, our implementation showed two drawbacks:
The more flexible quantization flow of TVM introduces annotation nodes into the code generation, which prevent the solution from reaching higher efficiency.
It is currently not possible to connect the BYOC flow with the micro-TVM runtime that is also still under development.
This prevents the usage of TVM on (heterogeneous) embedded devices for TinyML applications, however, it can already be utilized during the hardware development
to evaluate the performance of prototypes with real-world test cases.
%This prevents its use in conjunction with microcontrollers and Cortex-M CPUs.
%An observation during the testing was that the coarse-grained ISAs of the accelerator is hard to adapt to new layer and operation types.
%While these kind of ISAs achieves good performance results by using extensive pipelines, they are inflexible. 
%This can be a limitation for hardware solutions that are supposed to be sold for longer time periods, as they are at the risk to quickly become obsolete, if they cannot be adapted to new layer and function types efficiently.
%This is especially problematic due to the fast innovation speed of the deep learning research community.
%
%\begin{itemize}
%  \item reusing open source project to target custom accelerator possible
%  \item but still a lot of open points
%  \item flexible quantization flows are important on none GP hardware
%  \item TVM does not have a flow between BYOC and µTVM runtime yet
%  \item ISA for deep learning accelerators are an interesting research field
%  \item flexibility is necessary due to fast development of AI research
%\end{itemize}

%\clearpage
\appendix
\section{Performance of Hardware Variants}
As the accelerator was simulated with different software flow versions, ISA variations and memory configurations, multiple performance estimations have been collected.
Each of the three test scenarios from section \ref{eval} has been tested with 64 or 128 PEs in the compute module of the accelerator and 512\,KiB,
1\,MiB (only for the last scenario) and 256\,MiB for the SRAM memory on the device. An additional parameter for the simulation was the hardware parallelization
of the workload, which was either input or output parallel. The input parallel configuration would split the workload along the channel dimension of the 
input feature map, while the output parallel version would parallelize the workload along the output feature map's channel axis.\\
The bars of the diagrams are segmented according to the cycles spent on each supported subgraph type. In the test case for MobileNetV1 only up to three different
subgraph types could be accelerated:
\begin{itemize}
 \item{CONV subgraph:\\
	The CONV (Convolutional) subgraphs contain the Conv2D layers of the network, which can be mapped to the accelerator.}
 \item{REQUANT subgraph:\\
	The REQUANT subgraphs contain the requantization operations that are executed in-between layers to convert the 32-bit output of the
	previous layer back to an 8-bit format.}
\item{DEPTH subgraph:\\
	The DEPTH (Depthwise Convolution) subgraph is only present in the second and third test scenario, as the first did not offer support for
	the depthwise convolutional layer that is contained in it.}
\end{itemize}

\begin{figure}
     \centering
     \begin{subfigure}[h]{\linewidth}
         \centering
         \includegraphics[width=0.9\linewidth]{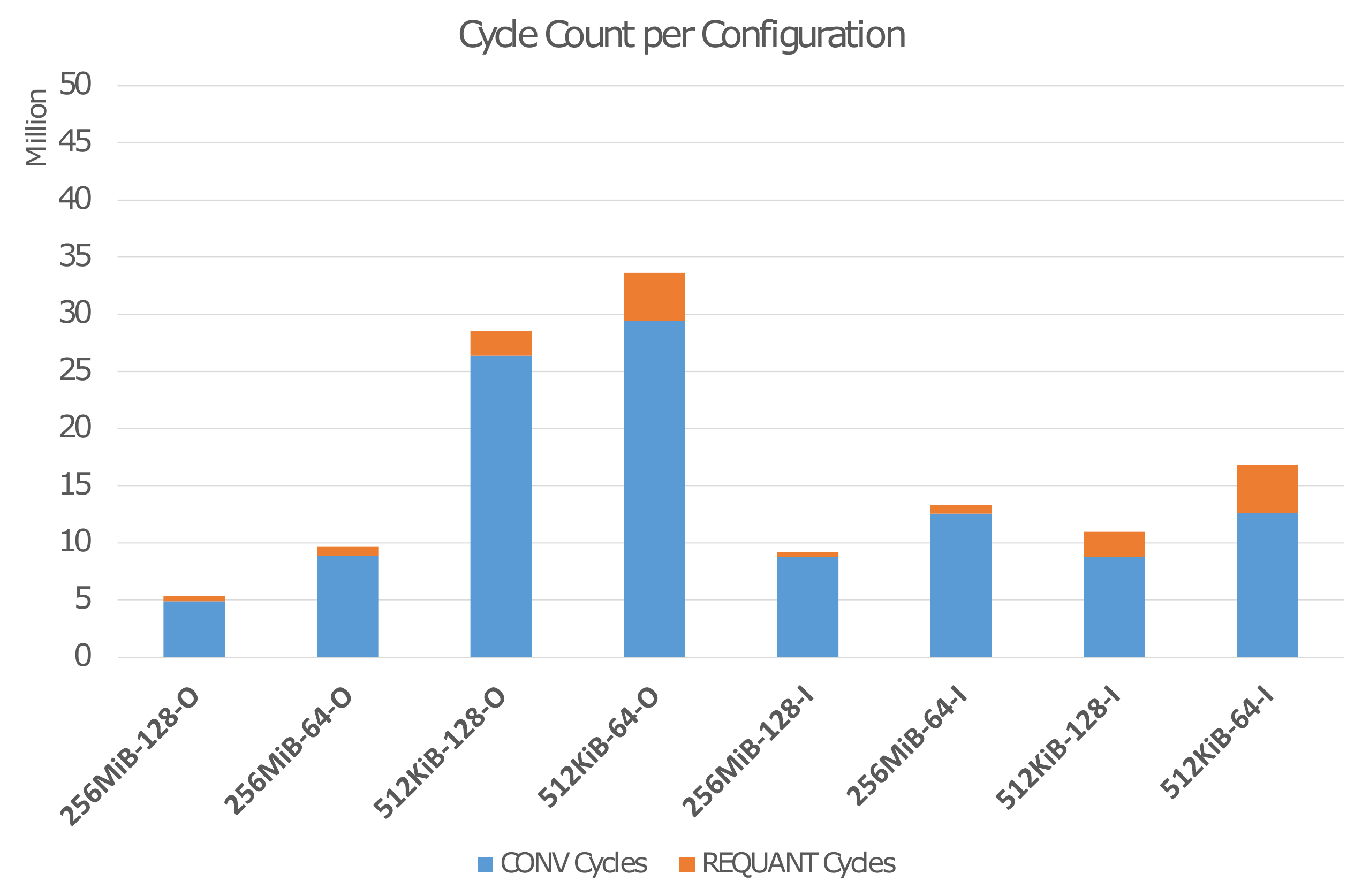}
         \caption{The total cycle count for the accelerated operations of MobileNet without support for depthwise convolutions across testing configurations.}
         \Description{Bar chart, showing estimated cycle count for the scenario without deptwhise convolution support.
256 MiB ASIP memory with 128 PEs, output parallel processing: CONV layers required 4.9 million cycles, requantizations about 434,000,
256 MiB ASIP memory with 64 PEs, output parallel processing: CONV layers required 8.9 million cycles, requantizations about 749,000
512 KiB ASIP memory with 128 PEs, output parallel processing: CONV layers required 26.4 million cycles, requantizations about 2.17 million
512 KiB ASIP memory with 64 PEs, output parallel processing: CONV layers required 29.4 million cycles, requantizations about 4.2 million
256 MiB ASIP memory with 128 PEs, input parallel processing: CONV layers required 8.8 million cycles, requantizations about 434,000
256 MiB ASIP memory with 64 PEs, input parallel processing: CONV layers required 12.6 million cycles, requantizations about 749,000
512 KiB ASIP memory with 128 PEs, input parallel processing: CONV layers required 8.8 million cycles, requantizations about 2.17 million
512 KiB ASIP memory with 64 PEs, input parallel processing: CONV layers required 12.6 million cycles, requantizations about 4.2 million}
         \label{fig:no_depth}
     \end{subfigure}
     \begin{subfigure}[h]{\linewidth}
         \centering
         \includegraphics[width=0.9\linewidth]{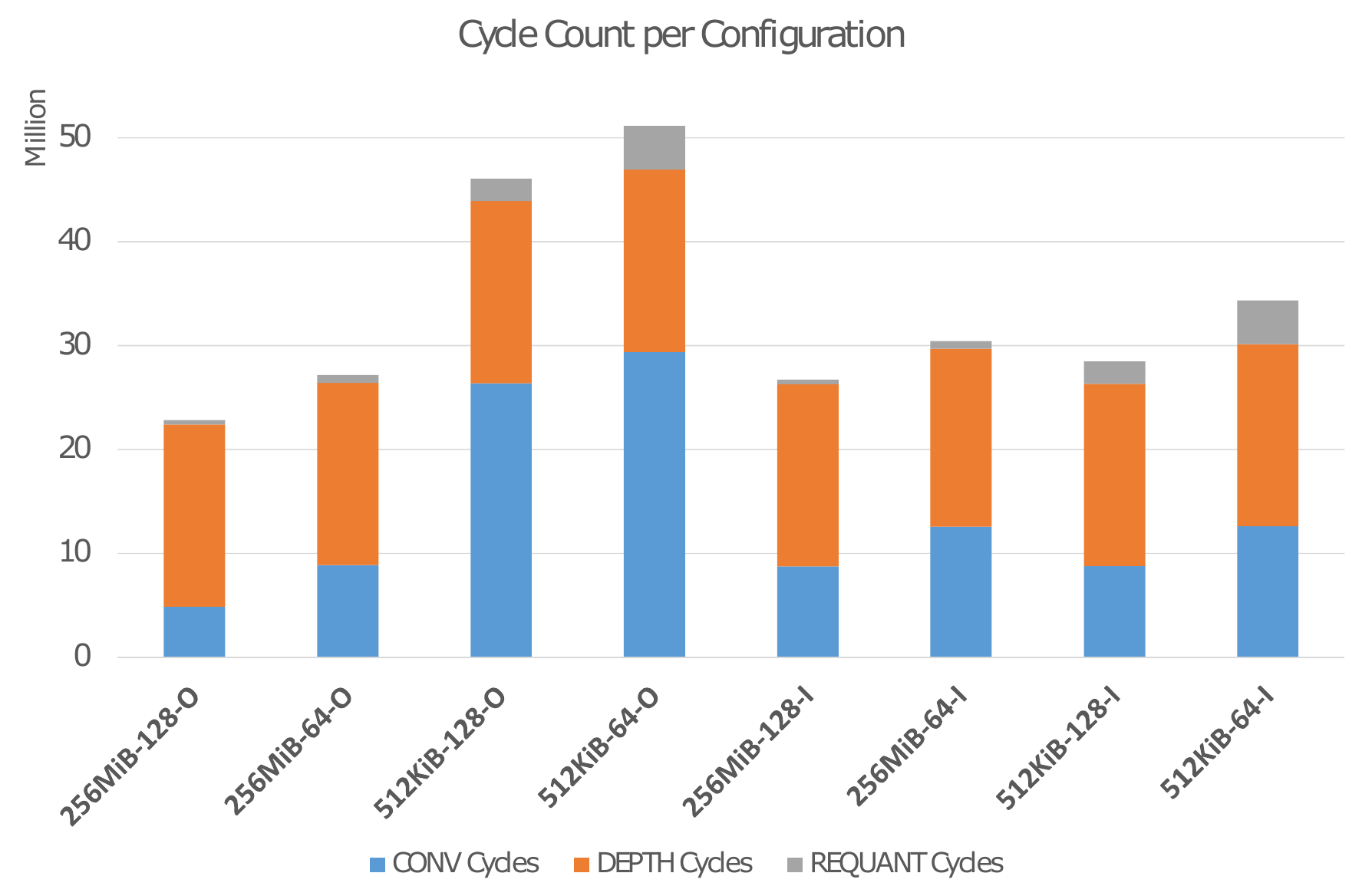}
         \caption{The total cycle count for the accelerated operations of MobileNet with software emulated support for depthwise convolutions across testing configurations.}
	\caption{The total cycle count for the accelerated operations of MobileNet with software-only support for depthwise convolutions across testing configurations.}
         \Description{Bar chart, showing estimated cycle count for the scenario without deptwhise convolution support.
256 MiB ASIP memory with 128 PEs, output parallel processing: CONV layers required 4.9 million cycles, requantizations about 434,000,
256 MiB ASIP memory with 64 PEs, output parallel processing: CONV layers required 8.9 million cycles, requantizations about 749,000
512 KiB ASIP memory with 128 PEs, output parallel processing: CONV layers required 26.4 million cycles, requantizations about 2.17 million
512 KiB ASIP memory with 64 PEs, output parallel processing: CONV layers required 29.4 million cycles, requantizations about 4.2 million
256 MiB ASIP memory with 128 PEs, input parallel processing: CONV layers required 8.8 million cycles, requantizations about 434,000
256 MiB ASIP memory with 64 PEs, input parallel processing: CONV layers required 12.6 million cycles, requantizations about 749,000
512 KiB ASIP memory with 128 PEs, input parallel processing: CONV layers required 8.8 million cycles, requantizations about 2.17 million
512 KiB ASIP memory with 64 PEs, input parallel processing: CONV layers required 12.6 million cycles, requantizations about 4.2 million.
The depthwise convolutional subgraphs always require 17.5 million cycles, across all hardware configurations.}
         \label{fig:sw_depth}
     \end{subfigure}
     \begin{subfigure}[h]{\linewidth}
         \centering
         \includegraphics[width=0.9\linewidth]{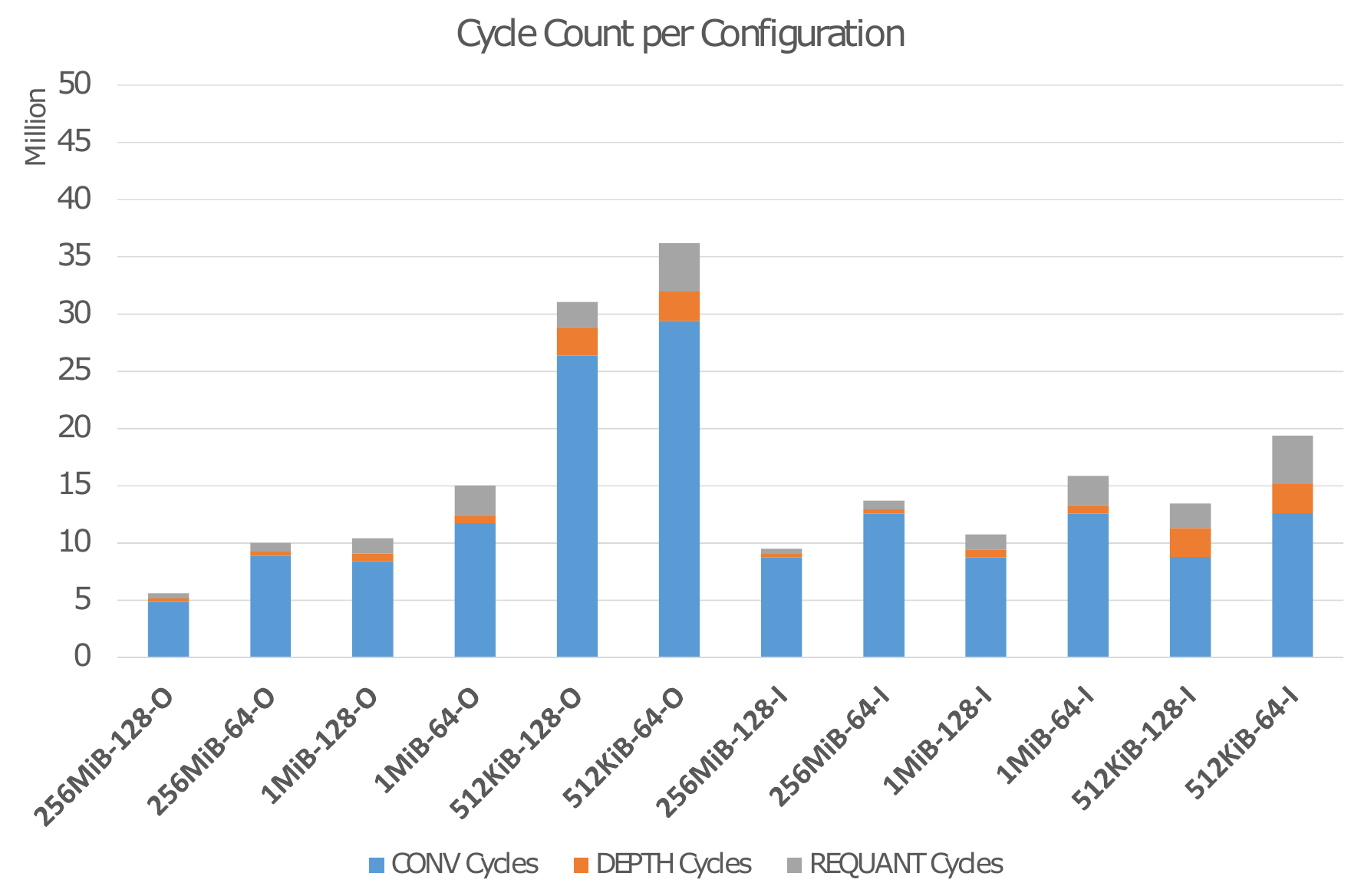}
         \caption{The total cycle count for the accelerated operations of MobileNet with hardware support for depthwise convolutions across testing configurations.}
\caption{The total cycle count for the accelerated operations of MobileNet without support for depthwise convolutions across testing configurations.}
         \Description{Bar chart, showing estimated cycle count for the scenario without deptwhise convolution support.
256 MiB ASIP memory with 128 PEs, output parallel processing: CONV layers required 4.9 million cycles, DEPTH takes 308,000, requantizations about 434,000,
256 MiB ASIP memory with 64 PEs, output parallel processing: CONV layers required 8.9 million cycles, DEPTH takes 401,000, requantizations about 749,000
1 MiB ASIP memory with 128 PEs, output parallel processing: CONV layers required 8.4 million cycles, DEPTH takes 677,000, requantizations about 434,000,
1 MiB ASIP memory with 64 PEs, output parallel processing: CONV layers required 11.7 million cycles, DEPTH takes 742,000, requantizations about 749,000
512 KiB ASIP memory with 128 PEs, output parallel processing: CONV layers required 26.4 million cycles, DEPTH takes 2.5 million, requantizations about 2.17 million
512 KiB ASIP memory with 64 PEs, output parallel processing: CONV layers required 29.4 million cycles, DEPTH takes 2.6 million, requantizations about 4.2 million
256 MiB ASIP memory with 128 PEs, input parallel processing: CONV layers required 8.8 million cycles,  DEPTH takes 308,000, requantizations about 434,000
256 MiB ASIP memory with 64 PEs, input parallel processing: CONV layers required 12.6 million cycles,  DEPTH takes 401,000, requantizations about 749,000
1 MiB ASIP memory with 128 PEs, input parallel processing: CONV layers required 8.8 million cycles, DEPTH takes 677,000, requantizations about 434,000
1 MiB ASIP memory with 64 PEs, input parallel processing: CONV layers required 12.6 million cycles, DEPTH takes 742,000, requantizations about 749,000
512 KiB ASIP memory with 128 PEs, input parallel processing: CONV layers required 8.8 million cycles, DEPTH takes 2.5 million, requantizations about 2.17 million
512 KiB ASIP memory with 64 PEs, input parallel processing: CONV layers required 12.6 million cycles, DEPTH takes 2.6 million, requantizations about 4.2 million}
         \label{fig:hw_depth}
     \end{subfigure}
        \caption{The first part of the label describes the size of the on-device memory, the second value represents the number of processing elements in the compute unit and the last letter stands either for an input (I) or output (O) parallel hardware implementation.}
        \label{fig:depth}
\end{figure}

\section{Target Accelerator}
To be more flexible a simulator has been used instead of a real hardware target as it came with several advantages, including 
a more flexible instruction set, easier configuration of hardware parameters. However, the simulator was still based on the typical 
concepts of other deep learning inference accelerators like the NVDLA including the coarse-grained ISA and the usage 
of highly optimized and separated hardware blocks for the different kinds of operations.\\
Its instruction set architecture was specifically designed for deep learning operations, and features dedicated instructions
for typical layer operations as well as common patterns like the combination of convolutional layers with activation functions and requantization operations,
which would be mapped to separate pipelines, that would combine multiple different functional units. Most instructions only support the 
processing of 8-bit integer data. A small subset of support instructions, like element-wise additions and shifts can also be executed on 32-bit data.\\
As most layers require a vast number of additional configuration parameters, which would not fit into a reasonably sized instruction, most 
of these parameters are stored in a separate register file and updated from the system CPU. This has the additional benefit that these values
can stay unchanged over multiple operations, without having to submit the same values multiple times, improving the energy efficiency of the solution.

\begin{table}[ht]
 \begin{center}
 \caption[Instruction Set]{Examples of instructions that are part of the accelerator ISA.}
 \label{tab:targets}
\bgroup
\def\arraystretch{1.2}
\resizebox{\linewidth}{!}{
\begin{tabular}{| l || p{3cm} |  c |  c |}
 \hline
Mnemonic                   & Description                                        & Input Type      & Output Type   \\
 \hline \hline
  \multicolumn{4}{|c|}{Compute Layer Instructions} \\
 \hline \hline
OP\_CONV                 & standard Conv2D, optional combined with requantization & int8                  & int32 or int8    \\
 \hline
OP\_CONV\_RELU     & Conv2D + ReLU                                                                   & int8                  & int8                  \\
 \hline
OP\_DEPTH\_CONV  & depthwise Conv2D, optional combined with requantization & int8                  & int32 or int8    \\
 \hline
OP\_MAT\_MUL        & Matrix Mulitplication                                                               & int8                  & int32                \\
 \hline \hline
  \multicolumn{4}{|c|}{Post-Processing Instructions} \\
 \hline \hline
OP\_ACT\_RELU      & ReLU Activation                                                                      & int8                  &  int8                 \\
 \hline
OP\_ACT\_LRELU    & Leaky ReLU Activation                                                            & int8                  &  int8                 \\
 \hline
OP\_POOL               & Pooling Layer                                                                          & int8                  &  int8                 \\
 \hline \hline
  \multicolumn{4}{|c|}{Elementwise Instructions} \\
 \hline \hline
OP\_E\_ABS            &  calculates absolute values of single input tensor                            & int8                  &  int8                 \\
 \hline
OP\_C\_MIN            & compares two tensors of same shape, writes minimum values       & int8                  &  int8                 \\
 \hline
OP\_C\_MAX            & compares two tensors of same shape, writes maximum values     & int8                  &  int8                 \\
 \hline
OP\_E\_ADD            & adds two tensors of same shape elementwise                    & int8                  &  int8                 \\
 \hline
OP\_E32\_ADD        & adds two tensors of same shape elementwise                    & int32                &  int32                 \\
 \hline \hline
  \multicolumn{4}{|c|}{DMA Instructions} \\
 \hline \hline
OP\_DMA\_READ    &  loads data from system memory to accelerator SRAM       & -                       &  -                      \\
 \hline
OP\_DMA\_WRITE  &  writes data from accelerator SRAM to system memory      & -                       &  -                      \\
 \hline

\end{tabular}}
\egroup
\end{center}
\end{table}

%\section{Appendix}

%\printbibliography
\bibliographystyle{abbrvnat}
\bibliographystyle{ACM-Reference-Format}
\bibliography{bib/sources, bib/related, bib/tvm, bib/glow, bib/ngraph, bib/tf, bib/xla, bib/onnc}
\end{document}